\documentclass[preprint,floatfix] {revtex4} 
\newcommand{\rvec}{\mathrm {\mathbf {r}}} 
\newcommand{\xvec}{\mathrm {\mathbf {x}}} 
\newcommand{\bvec}{\mathrm {\mathbf {b}}} 
\newcommand{\Rvec}{\mathrm {\mathbf {R}}} 
\newcommand{\Cvec}{\mathrm {\mathbf {C}}} 
\newcommand{\Svec}{\mathrm {\mathbf {S}}} 
\newcommand{\Ivec}{\mathrm {\mathbf {I}}} 
 
\newcommand{\Fvec}{\mathrm {\mathbf {F}}} 
\newcommand{\Pvec}{\mathrm {\mathbf {P}}}

\usepackage{graphicx}
\usepackage{subfigure}
\usepackage{xcolor}
\usepackage{enumerate}
\usepackage{braket}
\usepackage{amsmath, bm}
\usepackage{hhline}

\usepackage{color, soul}
\definecolor{darkblue}{rgb}{0,0,0.5}
\setulcolor{darkblue}

\begin{document}

\title{Static polarizability and hyperpolarizability in atoms and molecules through a Cartesian-grid DFT}

\author{Tanmay Mandal, Abhisek Ghosal}
\author{Amlan K.~Roy}
\altaffiliation{Email: akroy@iiserkol.ac.in, akroy6k@gmail.com.}
\affiliation{Department of Chemical Sciences\\                                            
Indian Institute of Science Education and Research (IISER) Kolkata, 
Mohanpur-741246, Nadia, West Bengal, India}

\begin{abstract}
Static electric response properties of atoms and molecules are reported within the real-space Cartesian grid implementation
of pseudopotential Kohn-Sham (KS) density functional theory (DFT). A detailed systematic investigation is made for a 
representative set of atoms and molecules, through a number of properties like total ground-state electronic energies, 
permanent dipole moment ($\bm{\mu}$), static average dipole polarizability ($\overline{\alpha}$). This is further extended to
first-hyperpolarizability ($\bm{\beta}$) in molecules. It employs a recently developed non-uniform grid-optimization technique, 
with a suitably chosen fixed initial applied field. A simple variant of the finite-field method, using a rational function 
fit to the dipole moment with respect to electric field, is adopted. We make use of Labello-Ferreira-Kurtz (LFK) basis set, 
which has performed quite well in these scenarios. To assess the efficacy and feasibility, four XC functionals such as LDA, 
BLYP, PBE and LBVWN are chosen. Present results are compared with available literature (both theoretical and experimental) 
values, whenever possible. In all instances, they show excellent agreement with 
the respective atom-centered-grid results, very widely used in many quantum chemistry programs. This demonstrates a viable 
alternative towards accurate prediction of response properties of many-electron systems in Cartesian coordinate grid. 

\vspace{5mm}

\end{abstract}
\maketitle

\section{Introduction}
In past several decades, we have witnessed remarkable stride in \emph{ab initio} quantum mechanical methods for 
accurate computation of various properties of many-electron systems, ranging from atoms, molecules to complex extended 
systems. To a large extent, this has been possible due to rapid advances in mathematical formalism, smart 
numerical algorithm, coupled with the developments in computer architecture. Amongst different theoretical approaches, 
Kohn-Sham density functional theory (KS-DFT) \citep{hohenberg64, kohn65} has proved to be a highly effective, versatile
tool for structural and electronic characterization of such systems. Within this framework, a considerable number of 
elegant and computationally efficient schemes are available to tackle these complex systems. They cover a broad range of 
approximations regarding the exchange-correlation (XC) functionals, keeping in mind the chemical accuracy as far as 
practicable. Due to its favorable computational cost, KS-DFT calculations are now routinely used in physical, chemical, 
biological and materials sciences, making them the preferred workhorse of electronic structure calculations for ``real-world" 
materials \citep{guliamov07}. The progress in the subject has been reviewed in many elegant references \citep{becke14, 
jones15}.

The electric-dipole polarizability i.e., response of electric dipole to an external electric field, is a very 
important property of an atom, molecule or cluster \citep{maroulis06, champagne10}. Its accurate, efficient 
description plays an important role in understanding electro-magnetic field-matter interaction and inter-particle collision 
phenomena such as, second-order Stark effect, Rayleigh and Raman scattering, 
electron-detachment process \citep{ghazaly05} etc. On the other hand, synthesis and design of novel non-linear optical
materials and molecular assemblies having large non-linear optical coefficients, are particularly interesting from a 
technological point of view \citep{kummel06}. Theoretical advances and dramatic enhancements in computational power in 
recent years have made it possible to provide guiding principles in this direction. 

A number of different theoretical approaches have been proposed in literature to compute electric response properties in 
linear and non-linear regime within the KS-DFT rubric. Some of the notable ones are: coupled-perturbed Kohn-Sham (CPKS) 
method \citep{fournier90, colwell93}, linear-response time dependent DFT \citep{jansik05, helgaker12}, 
perturbative sum-over states expression over all dipole-allowed electronic transitions \citep{orr71, bishop94}, 
numerical method using Sternheimer approach \citep{talman12}, 
auxiliary DFT \citep{moreno08, espindola12}, non-iterative CPKS \citep{sophy08} as well as the fully numerical finite field 
(FF) method \citep{bishop87, maroulis88}. From a computational viewpoint, FF method is the least expensive 
as there is no requirement of analytical derivatives or excited-state information, and also easy to implement as only 
one-body Hamiltonian is perturbed by the applied field \citep{kurtz90}. That is why FF method is usually considered at 
a first glance to justify the suitability and viability of a newly proposed quantum chemistry method for computation of 
electric response \citep{bulat05, wouters12, wergifosse14, wouters16}. The latter in a system is 
governed by its static dipole polarizability $\left(\alpha_{ij}=\frac{\partial{\mu_i}}
{\partial F_j}\right)$ as well as first $\left(\beta_{ijk}= \frac{\partial^2 \mu_i} {\partial F_j \partial F_k}\right)$ and 
second $\left(\gamma_{ijkl}=\frac{\partial^3 \mu_i}{\partial F_j \partial F_k \partial F_l}\right)$ hyperpolarizability, 
where $i,j$ denote the Cartesian directions $x,y,z$. Pertinent coefficients are then obtained from a Taylor expansion 
of the dipole moment, 
\begin{equation}
\mu_i(\Fvec) = \mu_i(0) + \sum_{j=1}^{3} \alpha_{ij} F_j + \frac{1}{2} \sum_{j,k=1}^{3} \beta_{ijk} F_j F_k + \cdots ~.
\end{equation}
where $\mu_i(\Fvec)=\sum_{m n} P_{m n}(\Fvec) \langle m |r_i|n \rangle + \sum_{\Rvec_{nuclei}} Z_{\Rvec} R_i$, 
denotes molecular dipole moment and $P_{m n}(\Fvec)$ the $(m,n)$th element of perturbed density matrix $\Pvec$ in presence of 
$-\rvec \cdot \Fvec$ perturbation, in dipole approximation. Alternatively one may also represent it in terms of 
field-induced energy, and both definitions are equivalent according to Hellmann-Feynman theorem. 
Now a vital factor in FF method is its numerical accuracy i.e, the system's dipole moment is calculated in presence of 
applied electric field, and the appropriate derivatives are approximated by finite differences. 
Hence the strength of applied field should be chosen carefully such that (i) it is sufficiently large to overcome 
finite-precision artifacts for a meaningful evaluation of necessary finite differences, especially in the non-linear regime for 
hyperpolarizabilities (ii) it must also be small enough so that the contributions from remaining higher-order derivatives 
become negligible for one particular coefficient. It is well documented that this effect is quite pronounced, in particular 
for higher-order derivatives \citep{bishop85}, and to optimize it, some novel techniques, \emph{viz.,} polynomial or 
rational-function-based FF methods (with or without extrapolation), were proposed in the literature \citep{
mohammed17, patel17}. 

A fundamental problem in a given DFT calculation lies in the correct approximation of 
hitherto unknown and difficult XC functional, regarding its proper asymptotic behavior at long-range as long as spurious 
self-interaction correction \citep{perdew81} which produces considerable error in these properties. In general, approximate 
local or semi-local functionals pose rather serious limitations regarding these issues. On the other hand, hybrid functionals 
(such as B3LYP or CAM-B3LYP) and their asymptotic-corrected variants were reported quite successfully in literature. Usually 
the obtained results are found to be as accurate as those from correlated wave-function based approaches \citep{castet12a}. 
This work employs following four XC functionals, namely, (i) local density approximation (LDA) using 
Vosko-Wilk-Nusair (VWN) \citep{vosko80} correlation, (ii) BLYP (Becke \cite{becke88} exchange with Lee-Yang-Parr (LYP) 
\cite{lee88} correlation) (iii) Perdew-Burke-Ernzerhof (PBE) \citep{perdew96} (iv) asymptotically corrected van Leeuwen-Baerends 
(LB94) \citep{leeuwen94} exchange in conjunction with VWN correlation. 
  
Recently KS-DFT was implemented in Cartesian coordinate grid (CCG) \cite{roy08, roy08a, roy11, ghosal16}, offering accurate 
results on atoms, molecules within an LCAO-MO ansatz. For a decent number of systems, several electronic 
properties including energy, {\boldmath$\mu$}, as well as its derivatives {\boldmath$\alpha, \beta$} were correctly 
presented. The electric responses were provided for three diatomic hydrides (HCl, HBr, HI) by including a 
coupling term (interaction of applied electric field and matter) in the respective one-body KS Hamiltonian \citep{ghosal18}. 
Through an accurate representation of FF formalism in real-space grid, corresponding results were compared and contrasted with 
the widely used, popular atom-centered grid ones (two were virtually indistinguishable) within a pseudopotential approximation. 
A detailed analysis of grid convergence and its impact on response was considered by taking four XC functionals, In order to 
broaden its scope of applicability, here we extend this to a larger set of atoms, molecules, which is the primary 
motivation of this communication. A secondary objective is to engage and ascertain the suitability of a recently proposed rational 
function approximation \cite{patel17} for their estimation. Note that in previous work, this was done through a Taylor FF approach. 

It is well acknowledged that the success of such calculations crucially depends on choice of an appropriate good-quality basis set. 
In other words, it should accurately describe the response of electrons to an external perturbation; an essential requisite being 
that it preferably includes polarization and diffused functions \cite{miadokova97}. A good candidate that satisfies these 
criteria is the Labello-Ferreira-Kurtz (LFK) basis \cite{labello05} utilized in our previous work \citep{ghosal18}, which has 
been found to be quite successful in both non-relativistic and relativistic domain. Current results are compared with available 
references (both theoretical and experimental), whenever possible, to gauge its performance. The article is organized as 
follows. Section II gives principal aspects of the method along with required computational details. Results are discussed
in Sec.~III, while the future and prospect is commented in Sec.~IV. 

\section{Methodology and computational aspects}
The single-particle KS equation of a many-electron system under the influence of pseudopotential can be written as (atomic 
unit employed unless stated otherwise), 

\begin{equation}
\bigg[ -\frac{1}{2} \nabla^2 + v^{p}_{ion}(\rvec) + v_{ext}(\rvec) + v_{h}[\rho(\rvec)] + v_{xc}[\rho(\rvec)] \bigg]
 \psi_i^{\sigma}(\rvec) = \epsilon_i \psi_i^{\sigma}(\rvec),
\end{equation}
where $v^{p}_{ion}$ denotes ionic pseudopotential, expressed as in the following,  
\begin{equation}
v^p_{ion}(\rvec) = \sum_{\Rvec_a} v^{p}_{ion,a} (\rvec-\Rvec_a).
\end{equation}
Here $v^{p}_{ion,a}$ represents ion-core pseudopotential associated with atom A, located at $\Rvec_a$. In presence of 
applied electric field, $v_{ext}(\rvec)$ contains contributions from both nuclear and field effects. The classical 
Coulomb (Hartree) potential, $v_{h}[\rho(\rvec)]$ describes electrostatic interaction amongst valence electrons 
whereas $v_{xc}[\rho(\rvec)]$ signifies the non-classical XC potential, and $\{ \psi^{\sigma}_{i},\sigma= 
\uparrow \mathrm{or} \downarrow \mathrm{spin}\}$ corresponds to a set of $N$ occupied orthonormal spin 
molecular orbitals (MO). In LCAO-MO approximation, the coefficients for expansion of MOs satisfy a set of equations 
analogous to that in Hartree-Fock theory,  
\begin{equation}
\sum_{n} F_{m n}^{\sigma} C_{n i}^{\sigma} = \epsilon_{i}^{\sigma} \sum_{n}^{\sigma} S_{m n} 
C_{ni}^{\sigma},
\end{equation}
with the orthonormality condition, $(\Cvec^{\sigma})^{\dagger} \Svec \Cvec^{\sigma} = \Ivec$. Here $\Cvec^{\sigma}$ contains 
respective MO coefficients $\{C_{n i}^{\sigma}\}$ for a given MO, $\psi_i^{\sigma}(\rvec)$, whereas $\Svec$ signifies the 
usual overlap matrix corresponding to elements $S_{m n}$ and $\bm{\epsilon}^{\sigma}$ refers to diagonal matrix of 
respective MO eigenvalues, $\{\epsilon_{i}^{\sigma}\}$. The KS matrix has elements $F_{m n}^{\sigma}$ 
partitioned as, 
\begin{equation}
F_{m n}^{\sigma} = H_{m n}^{\mathrm{core}} + J_{m n} + F_{m n}^{xc^{\sigma}}.
\end{equation} 
In this equation, $H_{m n}^{\mathrm{core}}$ contains all one-electron contributions and in presence of applied 
electric field this could be further separated as, $H_{m n}^{\mathrm{core}} = H_{m n}^{(0)} + F_i\langle m|r_i|n 
\rangle$, where $H_{m n}^{(0)}$ includes kinetic energy, nuclear-electron attraction and pseudopotential matrix 
elements, whereas the latter term signifies external electric field perturbation within dipole approximation. Furthermore, 
$F_i$ ($i=x,y,z$) corresponds to \textit{i}th component of field $\Fvec$, while $\langle m|r_i|n \rangle$ 
provides the dipole moment integral corresponding to length vector $\rvec$. Last two terms $J_{m n},F_{m n}^{xc\sigma}$ 
refer to usual contributions from classical Hartree and XC potentials respectively. 
 
Now all the relevant quantities like basis functions, electron densities, MOs as well as various two-electron potentials are 
directly set up on the $3$D CCG simulating a cubic box,
\begin{eqnarray}
r_{i}=r_{0}+(i-1)h_{r}, \quad i=1,2,3,....,N_{r}~, \quad r_{0}=-\frac{N_{r}h_{r}}{2}, \quad  r \in \{ x,y,z \},
\end{eqnarray}
where $h_{r}$ denotes grid spacing along each directions and $N_x, N_y, N_z$ signify total number of grid points along $x, 
y, z$ directions respectively. In case of non-uniform grid, we usually vary $N_{x}, N_y, N_{z}$ independently keeping the 
value of $h_{r}$ fixed. Thus, electron density $\rho(\rvec)$ in this grid may be simply written as (``g" implies the 
discretized grid),
\begin{equation}
\rho(\rvec_g) = \sum_{m,n} P_{m n } \chi_{m}(\rvec_g) \chi_{n}(\rvec_g). 
\end{equation}
A major concern in grid-based approach constitutes a correct, reliable estimation of two-electron KS matrix elements. In this work, 
they are calculated directly through straightforward numerical integration on grid as ($v_{hxc}$ refers to 
Coulomb and XC potential combined), 
\begin{equation}
 \langle \chi_{m}(\rvec_g)|v_{hxc}(\rvec_g)|\chi_{n}(\rvec_g) \rangle = h_x h_y h_z \sum_g
 \chi_{m}(\rvec_g) v_{hxc}(\rvec_g) \chi_{n}(\rvec_g).
 \end{equation}
The detailed construction of various potentials in grid has been well documented 
\citep{roy08, roy08a, roy11, ghosal16}.

As already mentioned, a critical problem of finite difference method for optical property calculation arises from the delicate 
nature of field--they satisfy a rather narrow range of field strengths. In order to alleviate this problem, we follow a recently 
published procedure \citep{patel17}, where the energy expression is fitted in terms of a rational function with respect to 
electric-field coefficients. An analogous fitting strategy for the induced dipole moment in terms of electric field is in invoked 
here, i.e., 
\begin{equation}
\bm{\mu}(\Fvec) = \frac{\ a+b\Fvec+c\Fvec^{2}+d\Fvec^{3}\cdots}{\ 1+B\Fvec+C\Fvec^{2}+D\Fvec^{3}\cdots}.
\end{equation}
where a,b,c,d,$\cdots$ and B,C,D,$\cdots$ are fitting coefficients. This is a generalized form of the Taylor expansion, as 
by setting all the denominator coefficients to zero, the latter expansion is recovered. This recipe provides a less sensitive 
(and hence consequently more effective) dependence on the electric field, as it increases latter's range. In FF 
technique, dipole moment of a given species needs to be computed at different field strengths, and its proper selection 
begins with a convenient choice of initial field strength ($F_0$), around which the field is distributed. This is achieved 
here via the following proposal \cite{patel17}, 
\begin{equation}
F_n = {x^{n}F_0}, \quad x=2^{\frac{p}{100}} 
\end{equation}
Here we choose their recommended value of $p$ (namely 50), corresponding to a geometry progression--this was reached on the basis
of an elaborate analysis of $\bm{\alpha}$ and $\bm{\gamma}$ for a set of 121 and $\bm{\beta}$ for 91 molecules. It is well 
established that the optimal value of $F_0$ plays a decisive role to ensure accuracy in these non-linear properties. On the 
basis of our recent work \cite{ghosal18}, where this aspect was discussed at length for $\bm{\beta}$, it is found to be 
appropriate to select an initial value of $10^{-2.5}$, which we exploit here. Next the \emph{optimal} (numerator and denominator 
degrees) form of rational function are taken from \citep{patel17} as, 
\begin{equation} 
\bm{\mu}(\Fvec) = \frac{\ a+b\Fvec+c\Fvec^{2}+d\Fvec^{3}}{\ 1+B\Fvec+C\Fvec^{2}},  
\end{equation} 
corresponding to four and three terms in the numerator and denominator respectively. Now putting the value of $\bm{\mu}$ at 
$\Fvec=0$ in Eq.~(11) gives rise to the trivial relation $ \bm{\mu}(0)=a$. The remaining unknown coefficients are determined using 
different $F_n$ values in this equation. For five unknown coefficients the minimum equations that can be constructed is six, because 
both $+F_n$ and $-F_n$ are used, with each $F_n$ producing two equations. By putting $\bm{\mu}(0)$ for $a$, the substituted 
equations may be recast in the matrix form $A\xvec=\bvec$ as, 
 \begin{equation}
 \begin{bmatrix} -F_0\bm{\mu}(F_0) & -F_0^2 \bm{\mu}(F_0) & F_0 & F_0^2 & F_0^3 \\
                  F_0\bm{\mu}(-F_0) & -F_0^2\bm{\mu}(-F_0) & -F_0 & F_0^2 & -F_0^3 \\
               -xF_0\bm{\mu}(xF_0) & -x^2F_0^2\bm{\mu}(xF_0) & xF_0 & x^2F_0^2 & x^3F_0^3 \\
                xF_0\bm{\mu}(-xF_0) & -x^2F_0^2\bm{\mu}(-xF_0) & -xF_0 & x^2F_0^2 & -x^3F_0^3 \\
            -x^2F_0^2\bm{\mu}(x^2F_0^2) & -x^4F_0^4\bm{\mu}(x^2F_0^2) & x^2F_0^2 & x^4F_0^4 & x^6F_0^6 \\
             x^2F_0^2\bm{\mu}(-x^2F_0^2) & -x^4F_0^4\bm{\mu}(-x^2F_0^2) & -x^2F_0^2 & x^4F_0^4 & -x^6F_0^6 \\ \end{bmatrix}
 \begin{bmatrix}  B \\ C \\ b \\ c \\ d \\ \end{bmatrix} 
 = \begin{bmatrix}
   \bm{\mu}(F_0) - \bm{\mu}(0) \\
   \bm{\mu}(-F_0) - \bm{\mu}(0) \\
   \bm{\mu}(xF_0) - \bm{\mu}(0) \\
   \bm{\mu}(-xF_0) - \bm{\mu}(0) \\
   \bm{\mu}(x^2F_0^2) - \bm{\mu}(0) \\
   \bm{\mu}(-x^2F_0^2) - \bm{\mu}(0) \\
   \end{bmatrix}
\end{equation}
Because both ($+$)ve and ($-$)ve fields are used, solution of this matrix equation is overdetermined, this can be solved by means of 
either least-squares method or by disregarding one of the equations. We adapt the second procedure, where the arbitrary elimination 
of any one of six equations was found to be valid for all the systems concerned. Finally, our desired properties can be determined 
by taking appropriate derivatives of Eq.~(11) at F=0 as,  
\begin{eqnarray}
\bm{\alpha} = \bm{\mu}'(0) = b-aB,  \nonumber \\ 
\bm{\beta} = \bm{\mu}''(0) = 2c-2bB-2aC-2aB^2.
\end{eqnarray}
   
A few practical aspects regarding the current compilation is in order. All calculations are performed using norm-conserving 
pseudopotential at the experimental geometries taken from NIST database \citep{johnson16}. The needed one-electron (excepting
the pseudopotential) integrals were generated by standard recursion relations \citep{obara86} using Cartesian Gaussian-type 
orbitals as primitive basis functions, whereas the latter matrix elements in Gaussian orbitals are imported from GAMESS 
\citep{schmidt93} package. The influence of spatial grid on 
energy components has been detailed in a previous article \citep{ghosal18} with respect to \emph{sparsity of grid} 
(governed by $N_x, N_y, N_z$) and \emph{grid spacing} (determined by $h_r$); and hence not repeated here. A simple grid 
optimization strategy has been maintained, which guarantees a grid accuracy of at least $5 \times 10^{-6}$ a.u., throughout, at 
a fixed $h_r = 0.3$. It was observed that the optimal non-uniform grid marginally varies from functional to functional. 
As done earlier, here also we employ the so-called Labello-Ferreira-Kurtz (LFK) basis set advocated in \citep{labello05} 
based on the procedure that incorporates diffuse and polarization functions in the well-known Sadlej \citep{sadlej92} 
basis--these are adopted from EMSL library \citep{feller96}. For efficient calculation of Hartree potential in real 
space, the standard discrete Fourier transform package FFTW3 \citep{fftw05} has been used. 

The properties are explored for four representative XC functionals, \emph{viz.,} LDA, BLYP, PBE, LBVWN. The middle 
two functionals were imported from density functional repository program \citep{repository}. 
The self-consistent convergence criteria imposed in this communication is slightly tighter than our earlier work 
\citep{roy08, roy08a, roy11, ghosal16}; this is to generate a more accurate perturbed density matrix. Convergence of following 
quantities was followed, \emph{viz.,} (i) orbital energy difference between two successive iterations (ii) absolute 
deviation in a density matrix element. They both were required to remain below a certain prescribed threshold set to 
$1\times 10^{-8}$ a.u.; this ensured that total energy maintained a convergence of at least this much. To accelerate the  
calculations, unperturbed (field-free) density matrix was used as trial input. The resulting generalized matrix-eigenvalue 
problem as well as Eq.~(12) was solved through standard LAPACK \citep{anderson99} routines. Scaling properties have been 
discussed elsewhere \cite{roy08}. The present calculations have been performed using an in-house computer 
program originally initiated by Roy in the references \cite{roy08, roy08a, roy11}, and subsequently extended \cite{ghosal16, 
ghosal18} by the two co-authors of this communication.

\section{Results and Discussion} 
We begin this section by categorizing two sets: first one for atoms containing 15 of them, labeled CCG-A and the second set
having 29 selected molecules, represented by CCG-M. Both sets contain open and close-shell electronic systems, with 
maximum number of total valence electrons being 32. The self-consistent field convergence was thoroughly checked with 
respect to all relevant parameters (like grid and field optimization) as delineated in previous section and elaborated in 
an earlier publication \cite{ghosal18}, both in absence/presence of the electric field. Since the converged properties thus 
obtained reproduce standard GAMESS results very well for all XC functionals available therein, these reference values are omitted 
henceforth. This matching agreement is expected and in keeping with the findings of \cite{ghosal18}, where this was demonstrated for 
three diatomic hydrides. A systematic investigation is carried out for these two sets through the following quantities, namely, 
non-relativistic ground-state total energy, $\bm{\mu}$, non-zero components of $\bm{\alpha}$, $\bm{\beta}$, along with 
the average or mean polarizability, $\overline{\alpha}$. 

\begingroup
\squeezetable
\begin{table}     
\caption{\label{tab:table1} Average static polarizability, $\overline{\alpha}$ for atoms in Set CCG-A, using FF KS method, 
for four XC functionals, along with the respective MSE and MAE's. All results in a.u. See text for details.}
\begin{ruledtabular}
\begin{tabular} {c|ccccc}
 & \multicolumn{5}{c}{Average static polarizability ($\overline{\alpha}$)}   \\
\cline{2-6} 
Atom & LDA & BLYP & PBE &LBVWN & Literature$^{a}$ \\
\cline{1-6}
Be($^1S$) & 44.49 & 43.43 & 43.10 & 40.81 & 37.79 \\ 
B($^2P$)  & 22.24 & 21.88 & 21.11 & 18.81 & 20.45 \\ 
C($^3P$)  & 13.68 & 13.55 & 13.50 & 10.68 & 11.88 \\ 
N($^4S$)  &  8.31 &  8.29 &  8.29 & 6.83 &  7.42 \\ 
O($^3S$)  &  5.62 &  5.48 &  5.47  & 4.21 & 5.41 \\ 
Mg($^1S$) & 76.91 & 75.02 & 74.23 & 70.51 & 71.53 \\ 
Al($^2P$) & 48.26 & 48.37 & 47.03 & 40.50 & 45.89 \\ 
Si($^3P$) & 37.50 & 37.89 & 36.23 & 35.13 & 36.31 \\ 
P($^4S$)  & 28.68 & 28.27 & 27.91 & 24.17 & 24.50 \\ 
S($^3P$)  & 21.75 & 20.91 & 20.54  & 18.28 &  19.57 \\ 
Cl($^2P$) & 16.25 & 16.51 & 15.73 & 13.84 & 14.71 \\ 
Ar($^1S$) & 12.14 & 12.02 & 11.88 & 10.43 & 11.07 \\ 
Ca($^1S$) & 165.32& 160.14 & 159.41  & 150.78 &153.86 \\ 
Kr($^1S$) & 18.21 & 18.14 & 17.86 & 15.59 & 16.77 \\ 
Xe($^1S$) & 28.67 & 28.42 & 28.04 & 25.02 & 27.29 \\ 
\cline{1-6}
MSE & $-$2.91 & $-$2.26 & $-$1.74 & 1.25 & -- \\
MAE &  2.91 & 2.26 & 1.74 & 1.65 & -- \\
\end{tabular}
\end{ruledtabular}
\begin{tabbing}
$^a$Theoretical values are from \cite{miller77}, as quoted in NIST database \citep{johnson16}.
\end{tabbing}
\end{table}
\endgroup

At first, the LDA, BLYP, PBE and LBVWN results for $\overline{\alpha}$ are provided for Set CCG-A in Table~I. As atoms contain 
inversion of symmetry, $\bm{\beta}$ vanishes, and the lowest non-vanishing non-linear coefficient is 
$\gamma$ (which is not considered in this work). For the sake of comparison, available theoretical values from NIST database 
\citep{johnson16} are quoted in last column. It reveals an interesting fact that, all three traditional functionals (LDA, BLYP, 
PBE) overestimate the references in the ranges of $3$-$11\%$, $1$-$9\%$ and $3$-$8\%$ respectively; however LBVWN offers 
underestimation in all cases (with exception for Be) by 5-9\%. Moreover we also provide the respective mean signed 
error (MSE) and mean absolute error (MAE) with respect to experimental result, for all the XC functionals, at the bottom 
of this table. It is observed that the first three functionals systematically overestimate the experimental values, while 
LBVWN shows underestimation except the lone case of Be. Furthermore the errors also diminish as one goes from columns 2-4, 
corresponding to LDA, BLYP and PBE, having same magnitudes for MSE and MAE. For LBVWN, the errors are lowest amongst all the 
four functionals; however these two deviations differ in magnitude considerably.  
Similar conclusions have also been drawn regarding the pattern behavior of these functionals 
for $\overline{\alpha}$ in \citep{vasiliev10}, where it was conjectured that, a significant improvement may be accomplished for 
$\bm{\alpha}$ by combining LB94 potential (in asymptotic region) with LDA (corrected for derivative discontinuity in the bulk 
region) exchange suitably. This leads to a more proper representation of exchange which approaches the experimental 
results quite closely \citep{casida98} at a lower level of computational cost than standard XC functionals. 

\begingroup
\squeezetable
\begin{table}     
\caption{\label{tab:table2} Ground-state energy and permanent static dipole moment (absolute) for all the molecules 
in CCG-M set, for four XC functionals. All results are in a.u. See text for details.}
\begin{ruledtabular}
\begin{tabular} {l|cccc|ccccc}
Molecule  & \multicolumn{4}{c}{$-\langle E \rangle$} &  \multicolumn{5}{c}{$\bm{\mu}$} \\
\cline{2-5} \cline{6-10}
   & LDA & BLYP & PBE & LBVWN & LDA & BLYP & PBE & LBVWN & Expt.$^{\dagger}$\\ 
\hline
F$_2$   & 47.91383 & 48.13149 & 48.12858 & 47.91426 & -- & -- & -- & -- & -- \\
Cl$_2$  & 29.65556 & 29.70811 & 29.40441 & 29.65489 & -- & -- & -- & -- & -- \\
Br$_2$  & 26.63157 & 26.66731 & 26.73636 & 26.63055 & -- & -- & -- & -- & -- \\
I$_2$   & 22.81646 & 22.82037 & 22.90226 & 22.81509 & -- & -- & -- & -- & -- \\
HF& 24.64387 &24.76034  &24.76552  &24.64197  & 0.70315 & 0.68988 & 0.69307 & 0.75623 & 0.71604 \\  
CO      & 21.49163 & 21.56884 & 21.59654 & 21.48968 & 0.09824 & 0.07539 & 0.09929 & 0.05770 & 0.04406\\
HCl     & 15.43342 & 15.47383 & 15.50513 & 15.43286 & 0.43825 & 0.42337 & 0.43420 & 0.45357 & 0.42490 \\
HBr     & 13.90305 & 13.93459 & 13.96864 & 13.90218 & 0.31610 & 0.29881 & 0.31207 & 0.30760 & 0.32536\\
HI      & 11.98212 & 11.99865 & 12.03800 & 11.98063 & 0.18224    & 0.16100    &0.17944    &0.11947  &0.17625 \\
H$_2$O& 17.06428 & 17.14538 & 17.16016 & 17.06321 & 0.71610 & 0.69956 & 0.70607 & 0.76583 & 0.7278\\ 
H$_2$S  & 11.25031 & 11.28181 & 11.31490 & 11.24899 & 0.44115 & 0.41985 & 0.43639 & 0.39553 & 0.38162 \\
H$_2$Se & 10.47641 & 10.50418 & 10.53809 & 10.47477 & 0.28488 & 0.25717 & 0.28047 & 0.19486 & 0.24668 \\
NH$_3$&11.61575 &11.67003 &11.69468  &11.61508  & 
0.57940 & 0.57091 & 0.57498 & 0.59063 & 0.57834 \\ 
PH$_3$  & 8.23604  & 8.26939  & 8.29887  & 8.23367  & 0.23137 & 0.19833 & 0.22442 & 0.11764 & 0.22818 \\
P$_4$   & 26.10747 & 26.10747 & 26.20419 & 26.10558 & -- & -- & -- & -- & -- \\
PCl$_3$ & 51.15601 & 51.20379 & 51.34330 & 51.15280 & 0.17708 & 0.21547 & 0.18162 & 0.35481 & 0.30687  \\ 
SiH$_3$Cl&20.49531 & 20.55085 & 20.60549 & 20.49152 & 0.50313 & 0.50656 & 0.49827 & 0.58014 & 0.51539 \\
SiH$_4$ & 6.16406 & 6.21318 & 6.23146 & 6.16040 & -- & -- & -- & -- & -- \\
CH$_3$Cl& 22.23814 & 22.28972 & 22.35264 & 22.23688&0.73076 & 0.73269 & 0.72914 & 0.71637 & 0.73571\\
CH$_2$Cl$_2$&36.51145 & 36.56834 & 36.66719 & 36.50986 & 0.63290 & 0.65394 & 0.63676 & 0.62398 & 0.62948 \\
CH$_3$Br& 20.70879 &20.75325 &20.81866 & 20.70720 & 0.71377 & 0.72486 & 0.71875 & 0.63353 & 0.71210 \\
CHCl$_3$& 50.78407 & 50.84088 & 50.97698 & 50.78216 & 0.42682 & 0.43710 & 0.42855 & 0.42664 & 0.39736 \\
CH$_4$   & 7.96412 & 7.96412 & 8.03747 & 7.96306 & -- & -- & -- & -- & --\\
CCl$_4$& 65.05251 & 65.10947 & 65.28327 &65.05034  
& -- & -- & -- & -- & --\\ 
C$_2$H$_2$   & 12.31925 & 12.35341 & 12.39988 & 12.31819 & -- & -- & -- & -- & --\\
C$_2$H$_4$ & 13.55522 & 13.55522 & 13.55522 & 13.55386 & -- & -- & -- & -- & -- \\
Si$_2$H$_6$& 11.19320 & 11.25222 & 11.29862 & 11.18691 & -- & -- & -- & -- & --\\                    
C$_3$H$_8$ & 21.58015 & 21.64097 &21.73504 & 21.57769 &0.04065 & 0.03925 & 0.03844 & 0.03102 & 0.03304 \\
C$_4$H$_6^{\ddagger}$ & 25.98404 & 26.03255 &26.14425 & 25.98181 & -- & -- & -- & -- & --\\                                                        
\end{tabular}
\end{ruledtabular}
\begin{tabbing}
${\dagger}$ For HCl and CHCl$_3$ Dielectric method \cite{nelson67}: For all others Microwave spectroscopy method \citep{nelson67}. \hspace{0.5cm}
$^{\ddagger}$(E)1,3-Butadiene. 
\end{tabbing}
\end{table}
\endgroup

Next in Table~II, we present ground-state energy and $\bm{\mu}$ (absolute value) for all 29 molecules in Set CCG-M. This 
covers diatomic to hexa-atomic systems containing both close and open shells--the equilibrium experimental geometries are 
taken from NIST computational chemistry database \cite{johnson16}. Note that these values pertain 
to only the electronic part; neither geometry relaxation in presence of electric field nor vibrational contribution is 
considered. In all occasions, total energies using BLYP and PBE XC functional are observed to be consistently lower than LDA 
and LBVWN. Keeping in mind the performance of results for various XC functionals in our earlier work \cite{roy08,
roy08a,roy11,ghosal16,ghosal18}, one can safely conclude that current total electronic energies for this data set are equally 
accurate and trustworthy. Several theoretical and experimental results are available in 
literature. In order to compare, we report only its non-zero components along with some selected experimental results. 
The computed value of zero components of $\bm{\mu}$ for non-polar molecules have been correctly reproduced in this calculation;
and henceforth not mentioned. In polar molecules also they show reasonably good agreement with experimental results. Leaving 
aside the two lone cases (CO plus PCl$_3$ for LDA, BLYP, PBE, and PH$_3$ for LBVWN), the overall maximum absolute deviation 
hovers around $13\%$ (for LDA, PBE) and $10\%$ (for BLYP, LBVWN) respectively from experimental counterparts. 

\begingroup
\squeezetable
\begin{table}     
\caption{\label{tab:table3} Average static polarizability, $\overline{\alpha}$ for CCG-M Set, using FF KS method 
for different XC functionals, along with MSE and MAE. All results are given in a.u. See text for details.}
\begin{ruledtabular}
\begin{tabular} {c|ccccc}
 Molecule  & \multicolumn{4}{c}{$\overline{\alpha}$} & Experiment$^{\dagger}$ \\
\cline{2-5} 
   & LDA & BLYP & PBE & LBVWN & (zero frequency) \\ 
\cline{1-6}
F$_2$       &  8.96 & 9.04  & 8.89  & 7.56  & 8.34\\
Cl$_2$      & 32.03 & 31.76 & 32.20 & 28.42 & 30.37\\
Br$_2$      & 46.60 & 45.89 & 45.60 & 41.65 & 43.7\\
I$_2$       & 72.68 & 71.54 & 71.07 & 66.10 & 70.36\\
CO          & 13.66 & 13.57 & 13.43 & 11.68 & 13.04\\
HF & 6.24 & 6.25& 6.15& 4.88& 5.09 \\ 
HCl         & 18.79 & 18.55 & 18.37 & 16.07 & 17.40\\
HBr         & 25.75 & 25.19 & 25.09 & 22.22 & 23.78\\
HI          & 37.97 & 37.09 & 36.96 & 33.35 & 35.30\\
H$_2$O      & 10.52 & 10.41 & 10.26 & 8.91  & 9.52\\
H$_2$S      & 26.45 & 26.01 & 25.76 & 22.65 & 24.68\\
H$_2$Se     & 31.74 & 31.29 & 30.85 & 27.00 & 29.68\\
NH$_3$      & 15.43 & 15.31 & 15.08 & 12.59 & 14.61\\
PH$_3$      & 32.49 & 32.69 & 29.44 & 26.29 & 30.90\\
P$_4$       & 95.75 & 94.90 & 93.85 & 88.44 & 91.70\\
PCl$_3$     & 74.13 & 73.34 & 72.86 & 67.03 & 70.30\\
SiH$_3$Cl   & 44.93 & 43.81 & 43.86 & 39.93 & 35.8\\
SiH$_4$     & 34.07 & 32.80 & 33.01 & 29.91 & 31.94\\
CH$_3$Cl    & 31.95 & 31.72 & 31.28 & 27.50 & 30.00\\
CH$_2$Cl$_2$& 47.48 & 46.94 & 46.55 & 42.46 & 44.89\\
CH$_3$Br    & 38.70 & 38.33 & 37.87 & 34.35 & 36.76\\
CHCl$_3$    & 60.12 & 59.52 & 58.98 & 52.02 & 56.22\\
CH$_4$      & 17.98 & 17.69 & 17.44 & 15.70 & 17.24\\
CCl$_4$     & 74.41 & 73.66 & 73.06 & 67.44 & 69.23\\
C$_2$H$_2$  & 24.46 & 24.78 & 24.02 & 20.86 & 22.67\\
C$_2$H$_4$  & 29.21 & 29.34 & 28.60 & 25.53 & 27.72\\
Si$_2$H$_6$ & 65.76 &63.32  &63.81  & 59.02 & 63.53 \\                    
C$_3$H$_8$  & 43.94 & 43.23 & 42.69 & 39.27 & 42.12 \\
C$_4$H$_6^{\ddagger}$  & 59.34 & 59.64 & 58.27 & 53.07 & 54.64 \\
\cline{1-6}
MAE & 2.59 & 1.95 & 1.61 & 2.33 &  -- \\ 
MSE & $-$2.59 & $-$1.93 & $-$1.51 & 2.18 & -- \\ 
\end{tabular}
\end{ruledtabular}
\begin{tabbing}
$^{\dagger}$For PH$_3$ and SiH$_3$Cl, dielectric permittivity method \cite{hohm13}; for all others, refractive index method \citep{hohm13}
 \hspace{0.5cm} \=
$^{\ddagger}$(E)1,3-Butadiene.
\end{tabbing}
\end{table}
\endgroup

Next we move towards the average polarizability, $\overline{\alpha}$ of molecules in Set CCG-M. Table~III reports these
at equilibrium geometries of NIST computational chemistry database \cite{johnson16}, for same four XC 
functionals of previous tables. To put things in perspective, we also quote corresponding experimental values of zero frequency 
(containing only electronic part) in column six, from the compilation of \citep{hohm13}. 
It uncovers that, all three traditional functionals (LDA, BLYP, PBE) overestimate 
experimental references in the range of $3$-$11\%$, $1$-$9\%$ and $3$-$8\%$ respectively; however, LBVWN separates out from them
by underestimating within $5$-$9\%$ (with exception of Be). {\color{red}We have also performed the MAE and MSE analysis, as given in 
the bottom. It follows similar kind of behavior regarding first three functionals (LDA, BLYP and PBE), in harmony with those of CCG-A set in 
Table~I. However, it is observed that the results using LBVWN shows much more error compared to BLYP and PBE in contrast to Table~I.}  

\begingroup
\squeezetable
\begin{table}     
\caption{\label{tab:table4} The components of first hyperpolarizability for some selected molecules from CCG-M set, 
for four XC functionals. All results are in a.u. See text for details.}
\begin{ruledtabular}
\begin{tabular} {l|cccc|cccc|cccc}
 Molecule & LDA & BLYP & PBE & LBVWN & LDA & BLYP & PBE & LBVWN& LDA & BLYP & PBE & LBVWN \\
\cline{1-13}
 & \multicolumn{4}{c}{$\beta_{xxz}$} & \multicolumn{4}{c}{$\beta_{yyz}$} & \multicolumn{4}{c}{$\beta_{zzz}$}\\
\cline{2-5} \cline{6-9} \cline{10-13}  
\cline{1-13}
 H$_2$S$^a$   & $-$12.41& $-$14.30 & $-$12.21 & $-$4.93 & 6.96    & 5.80     & 6.07   & 8.78   & $-$25.07& $-$27.54 & $-$24.92 & $-$7.31 \\ 
 H$_2$Se      & $-$23.65& $-$32.56 & $-$27.08 & $-$8.95 & 52.99   & 46.07    & 48.75  & 40.85  & 9.24    & 0.58     & 6.58     & 15.67 \\ 
 PH$_3$$^b$   & 6.07    & 4.56     & 6.38     & 5.19    & 6.07    & 4.56     & 6.39   & 5.19   & 20.73   & 5.63     & 14.79    & 6.12 \\ 
 SiH$_3$Cl    & 23.01   & 18.19    & 19.86    & 14.93   & 23.01   & 18.18    & 19.86  &14.93   & 58.69   & 53.77    & 55.75    & 46.66 \\ 
 CH$_3$Cl     & $-$6.64 & $-$8.82  & $-$8.44  &  0.34   & $-$6.65 & $-$8.81  &$-$8.43 & 0.33   & 21.24   & 18.87    &17.19     & 19.55 \\ 
 CH$_2$Cl$_2$ & 14.45   & 14.62    & 14.66    &  2.13   & $-$26.39& $-$26.68 &$-$25.51&$-$19.88& 23.99   & 24.12    &24.48     & 31.15 \\
CHCl$_3$$^{c}$& $-$19.42& $-$18.65 & $-$17.82 & $-$11.75& $-$18.49& 
$-$17.80 &$-$16.92&$-$11.42& 15.31   & 17.06    & 15.94    & 2.17 \\
H$_2$O$^d$& 18.73& 14.76 & 16.67 & 8.60 & 16.19 & 16.26 &
15.85 & 11.40 & 31.78   & 26.75    & 28.75    & 19.75 \\
NH$_3$$^e$& $-$13.97& $-$15.65 & $-$14.15 & $-$9.73 & $-$13.97 & 
$-$15.65 &$-$14.15 &$-$9.73 & $-$55.43   & $-$51.10 & $-$51.20 & 
$-$26.37 \\
\cline{1-13}
  & \multicolumn{4}{c}{$\beta_{xxy}$} & \multicolumn{4}{c}{$\beta_{yzz}$} & \multicolumn{4}{c}{$\beta_{yyy}$}\\
\cline{1-13}
C$_3$H$_8$&1.04&2.99&1.69&1.82&$-$25.01&$-$25.15&
$-$23.92&$-$14.13&$-$28.59&$-$26.24 &$-$26.30 &$-$11.06 \\
\cline{1-13}
  & \multicolumn{4}{c}{$\beta_{xyy}$} & \multicolumn{4}{c}{$\beta_{xzz}$} & \multicolumn{4}{c}{$\beta_{xxx}$}\\
\cline{1-13}
CH$_3$Br&42.62&45.80&4.84&21.94&42.59& 45.86&
44.13&21.94&17.24&18.72&21.43&5.40 \\
\cline{1-13}
  & \multicolumn{4}{c}{$\beta_{xxz}$=$\beta_{yyz}$} & \multicolumn{4}{c}{$\beta_{zzz}$}\\
\cline{2-5} \cline{6-9} 
 CO$^{f}$       & 8.55    & 7.88     & 8.42   & 3.44   & 31.83   & 30.29    & 29.80    & 21.48 \\ 
HF$^g$     & $-$3.59& $-$3.24&  $-$3.22 & 
$-$1.51 & $-$14.09  & $-$14.04& $-$13.52& $-$9.24 \\
 HCl$^{h}$  & 8.27    & 6.30     & 7.19   & 3.78   & 20.77   & 19.60    & 18.91    & 15.19 \\ 
 HBr$^{i}$  & 7.28    & 4.16     & 5.79   & 2.21   & 23.13   & 20.90    & 20.52    & 15.61 \\ 
 HI         & $-$3.32 & 1.19     &$-$1.80 & 2.39   & $-$16.48& $-$12.64 & $-$13.22 & $-$9.22 \\ 
\end{tabular}
\end{ruledtabular}
\begin{tabbing}

$^a$CCSD result in polarizability consistent Sadlej basis \citep{sekino93}: $\beta_{zzz}$=7.7, $\beta_{xxz}$=$-$1.2 and 
$\beta_{yyz}$=$-$11.7.  Experimental value of \\
$\langle \beta \rangle$ = $\sqrt(\sum_{i}\beta_{i}^2)$ = $-10.1$,$\beta_{i}=(1/3)\sum_{k}\beta_{ikk}$, obtained from \citep{sekino93}. \\

$^b$CCSD result in NLO-II basis \citep{paschoal14}: $\langle \beta \rangle$ = $\sqrt(\sum_{i}\beta_{i}^2)$= $-$18.5, 
$\beta_{i}=(1/3)\sum_{k}\beta_{ikk}.$ \\

$^c$CCSD-QRF result in d-aug-cc-pVDZ basis: $\beta_{HRS}=15.05$ and TDHF result in d-aug-cc-pVDZ basis: $\beta_{HRS}=10.02$ 
 \citep{wergifosse15}; \\
Experimental value of $\beta_{HRS} = -19.0$, obtained from Hyper-Rayleigh scattering experiment \citep{castet12};
{\color{red}$\beta_{HRS}=\sqrt{(\langle \beta_{ZZZ}^2\rangle + \langle \beta_{XXZ}^2\rangle)}$,} \\ 
{\color{red}coresponding to laboratory axes.} \\

$^d$CCSD result in polarizability consistent Sadlej basis \citep{sekino93}:  $\beta_{zzz}$=$-$7.3, $\beta_{xxz}$ = $-$2.0
and $\beta_{yyz}$=$-$10.8; Experimental value of \\
$\langle \beta \rangle = -22.0$, obtained from \citep{maroulis91}. \\

$^e$CCSD result in polarizability consistent Sadlej basis \citep{sekino93}:  $\beta_{zzz}$=$-$26.4 and $\beta_{xxz}$ = 
$\beta_{yyz}$=$-$7.6; Experimental value of \\
$\langle \beta \rangle = -48.9$, obtained from \citep{maroulis92}. \\

$^{f}$CCSD result in polarizability consistent Sadlej basis \citep{sekino93}:  $\beta_{zzz}$=26.1 and $\beta_{xxz}$ = 
$\beta_{yyz}$=6.1. Experimental value of \\
$\langle \beta \rangle$ = $\sqrt(\sum_{i}\beta_{i}^2)$= 30.2, 
$\beta_{i}=(1/3)\sum_{k}\beta_{ikk},$ obtained from \citep{shelton94}. \\

$^g$CAS result in taug-cc-pVTZ basis \citep{bishop99}: $\beta_{xxz} = \beta_{yyz} = -1.2$, $\beta_{zzz}$ = $-$8.77. CCSD result in 
polarizability consistent Sadlej \\
basis \citep{sekino93}: $\beta_{xxz}$ = $\beta_{yyz}$ = $-$0.08, $\beta_{zzz}$ = $-$8.91; Experimental value of 
$\langle \beta \rangle = 11.0$, obtained from \citep{shelton94}.\\

$^h$CAS result in taug-cc-pVTZ basis \citep{bishop99}: $\beta_{xxz} = \beta_{yyz} = -0.31$, $\beta_{zzz}$ = $-$11.32. CCSD(T) result in KT1 
basis \citep{maroulis98}: \\
$\beta_{xxz}$ = $\beta_{yyz}$ = $-$0.2, $\beta_{zzz}$ = $-$10.7. Experimental value of $\langle \beta \rangle = 9.8$, obtained from 
\citep{dudley85}. \\

$^i$CAS result in taug-cc-pVTZ basis \citep{bishop99}: $\beta_{xxz}$=$\beta_{yyz}$ = 1.41, $\beta_{zzz}$= $-$11.13. CAS result in qaug-Sadlej 
basis \citep{fernandez98}: \\
$\beta_{xxz}$ = $\beta_{yyz}$ = $-$0.81, $\beta_{zzz}$ = 11.14. \\

\end{tabbing}
\end{table}
\endgroup

In the last part, Table~IV reports the non-zero components of $\bm{\beta}$ (using the $T$ convention) for some 
representative molecules (CO, HCl, HBr, 
HI, H$_2$S, H$_2$Se, PH$_3$, SiH$_3$Cl, CH$_3$Cl, CH$_2$Cl$_2$, CHCl$_3$, C$_3$H$_8$, CH$_3$Br, H$_2$O, NH$_3$, HF) 
of CCG-M set (for same four XC functionals) to extend the 
scope of applicability. The results for HCl, HI and HBr using celebrated FF method are already mentioned in our previous 
article \citep{ghosal16} and these remain nearly same in the present case too.  
Same field strength, as mentioned in Sec.~II has been employed irrespective of the molecular system and XC functional. 
It is clear that for a particular molecule the components differ significantly from functional to functional--sometimes even 
the signs vary. One such candidate is HI, where $\beta_{xxz}, \beta_{yyz}$ signs for LDA, PBE functionals are opposite from 
those of BLYP, PBE. In literature these properties have been pursued through a wide range experiments and theoretical calculations. As a 
matter of comparison, some selected high-level all-electron calculations (such as CCSD, CAS, CCSD(T)) in large basis sets 
(Sadlej, taug-cc-pVTZ, qaug-sadlej, NLO-II) are quoted, along with some selected experiments. Of course, as expected, for obvious reasons 
present results differ 
from these extended and elaborate results rather considerably. However this is aside the main 
focus of this work; here we wanted to extend the domain of CCG in the context of response properties. As evident from 
the foregoing discussion (along with the previous tables), this scheme very satisfactorily fulfills that objective, for a decent 
number of atoms, molecules. However, as well known, for better comparison with experiment and other sophisticated 
theoretical calculations, more suitable basis set as well as XC functionals having correct asymptotic potential at long range, 
would be necessary, which may be pursued in our future works. 
 
\section{Future and outlook}
We have analyzed the performance of CCG in the context of $\bm{\mu}, \bm{\alpha}$ for a set of 44 species (sets CCG-A and 
CCG-M) using first-principles pseudopotential DFT formalism combined with an optimized FF procedure. Additionally, $\bm{\beta}$ 
values were provided for ten molecules. This was realized through a recently prescribed rational function approximation of 
dipole moment in real-space grid within the LFK basis set. The suitability and viability of this simple yet effective scheme has 
been demonstrated for four different XC functionals. Comparison with existing theoretical and experimental results have been made, 
wherever possible. In all occasions, our present results show excellent agreement with the corresponding values from familiar 
atom-centered grid. For further accuracy, more precise basis set
as well as XC functionals need to be invoked. In some DFT works \cite{yanai04,limacher09}, significant improvements in these quantities 
have been reported through the use of popular CAMB3LYP functional; we would like to consider this in a forthcoming
communication. Moreover, it will also be interesting to study the dynamical hyper-polarizability within the rubric
of time-dependent DFT. To conclude, pseudopotential-CCG can offer fairly accurate and reliable results for electric response 
in atoms and molecules.

\section{Acknowledgement}
AG thanks UGC for a Senior Research Fellowship. TM very much appreciates a Junior Research fellowship from IISER Kolkata. 
Financial support from DST SERB, New Delhi, India (sanction order number EMR/2014/000838) is sincerely acknowledged. Constructive 
comments from anonymous referee have helped in improving the quality of this manuscript.

\end{document}